\documentclass[10pt,showpacs,twocolumn,nofootinbib,aps,prd]{revtex4-1}
\usepackage{amsmath,amssymb,bm,mathrsfs}
\usepackage{mathtools,mathptmx,array,slashed}
\usepackage{graphicx,graphics,subfigure}
\usepackage{ctable,multirow}
\usepackage{tabularx}
\usepackage[colorlinks=true,linkcolor=blue,citecolor=blue,urlcolor=blue]{hyperref}

\DeclareSymbolFont{symbols} {OMS}{cmsy}{m}{n}

\def\be{\begin{equation}}
\def\ee{\end{equation}}
\def\bea{\begin{eqnarray}}
\def\eea{\end{eqnarray}}
\def\ba{\begin{aligned}}
\def\ea{\end{aligned}}
\def\nn{\nonumber}

\begin{document}


\title{Universal thermodynamic topological classes of three-dimensional BTZ black holes}

\author{Ying Chen}

\author{Xiao-Dan Zhu}
\email{Corresponding author: zxdcwnu@163.com}

\author{Di Wu}
\email{Corresponding author: wdcwnu@163.com}

\affiliation{School of Physics and Astronomy, China West Normal University, Nanchong, Sichuan 637002, People's Republic of China}

\date{\today}

\begin{abstract}
We establish a universal thermodynamic topological classification for three-dimensional static neutral Ba\~{n}ados-Teitelboim-Zanelli (BTZ), charged BTZ, and rotating BTZ black holes. We demonstrate that in all three cases (static neutral BTZ, charged BTZ, and rotating BTZ black holes), both the innermost small black hole states and the outermost large black hole states exhibit stable thermodynamic behavior. In the low-temperature limit, all three cases exhibit a thermodynamically stable small black hole state. Conversely, in the high-temperature limit, each system admits a thermodynamically stable large black hole state. Through this analysis, we have rigorously shown that static neutral, charged, and rotating BTZ black holes are consistently classified within the $W^{1+}$ category. Our results demonstrate that neither the charge parameter nor the rotation parameter exerts significant influence on the universal thermodynamic topological classification of three-dimensional static neutral BTZ black holes. This reveals a fundamental dichotomy: while angular momentum and electric charge dominate the thermodynamic topology of four-dimensional static black holes, their effects become negligible in the three-dimensional static BTZ case, highlighting a dimension-driven divergence in black hole thermodynamic behavior.
\end{abstract}

\maketitle


\section{Introduction}
Black hole thermodynamics explores the profound interplay between thermodynamic laws and the fundamental properties of black holes, bridging classical gravity and quantum mechanics. However, to date, uncovering universal thermodynamic properties of black holes remains a challenging endeavor. Progress in this field has been driven by topological methods \cite{PRL129-191101,PRD110-L081501,PRD111-L061501}, wherein black hole solutions are framed as topological defects in the thermodynamic parameter space.\footnote{Topological approaches also serve as a powerful tool for studying light rings \cite{PRL119-251102,PRL124-181101,
PRD102-064039,PRD103-104031,PLB858-139052,2412.18083} and timelike circular orbits \cite{PRD107-064006,JCAP0723049,2406.13270}.} Early studies classified these defects into three categories based on topological invariants \cite{PRL129-191101}, a classification later refined in Ref. \cite{PRD110-L081501} to group solutions into four broader classes according to their asymptotic thermodynamic behaviors. Building on this framework, Ref. \cite{PRD111-L061501} identified a novel topological class and two distinct subclasses, significantly expanding the taxonomy. \footnote{For recent developments and applications, please see Refs. \cite{PRD107-024024,PRD107-064023,PRD107-084002,EPJC83-365,EPJC83-589,PRD108-084041,JHEP0624213,
PLB856-138919,PDU46-101617,2409.11666,PLB860-139163}.} A brief review of the thermodynamic topological method is presented below.

The topological properties of black hole thermodynamics are fundamentally rooted in the concept of generalized off-shell free energy \cite{PRD33-2092}. In this framework, a black hole with mass $M$ and entropy $S$ is enclosed within a cavity maintained at a fixed temperature $1/\tau$. This setup gives rise to a generalized free energy defined as
\be
\mathcal{F} = M -\frac{S}{\tau}, \label{F}
\ee
which can also be derived from the gravitational path integral \cite{PRD106-106015}. Notably, the free energy reduces to an on-shell quantity only when $\tau = \beta = 1/T$, where $T$ represents the Hawking temperature of the black hole.

To further analyze the topological structure, an additional parameter $\Theta \in (0, \pi)$ is introduced, enabling the definition of a two-component vector field:
\be
\phi = \big(\phi^{r_h}, \phi^\Theta\big) = \Big(\frac{\partial \mathcal{F}}{\partial r_h}, -\frac{\cos\Theta}{\sin^2\Theta}\Big) \, .
\ee
Here, the condition $\phi^{r_h} = 0$ identifies black hole states as zero points (or topological defects) of the vector field. By applying Duan's $\phi$-mapping topological current theory \cite{SS9-1072}, each zero point-or equivalently, each black hole state-can be assigned a topological charge, known as the winding number $w$ \cite{PRL129-191101}.

From a local perspective, black hole states exhibit distinct topological characteristics: locally stable states are associated with a positive winding number $w = +1$, while locally unstable states are characterized by a negative winding number $w = -1$. On a global scale, the sum of all local winding numbers $w_i$ for a given class of black holes yields the global topological number $W$. This global quantity provides a robust basis for the topological classification of black holes, offering a unified perspective on their thermodynamic stability and phase structure. Notably, the three-dimensional Ba\~{n}ados-Teitelboim-Zanelli (BTZ) black hole has established itself as a pivotal framework for addressing fundamental questions in quantum gravity, thermodynamics, and high-energy theoretical physics. Investigating their universal thermodynamic topological classes thus becomes imperative, serving as the central motivation for this work.

In this paper, we investigate the universal thermodynamic topological classification of three distinct BTZ black holes: the static neutral BTZ black hole, the charged BTZ black hole, and the rotating BTZ black hole. We find that the static neutral BTZ, charged BTZ, and rotating BTZ black holes all belong to the class $W^{1+}$. The remaining part of this paper is organized as follows. In Sec. \ref{II}, we investigate the universal thermodynamic topological classification of the static neutral BTZ black hole, analyzing its state systematic ordering and universal thermodynamic behavior. Secs. \ref{III} and \ref{IV} extend this analysis to the charged BTZ black hole and the rotating BTZ black hole, respectively. Finally, we present our conclusions in Sec. \ref{V}.

\section{Static neutral BTZ black hole}\label{II}
In this section, we will investigate the universal thermodynamic topological class of the three-dimensional static neutral BTZ black hole, whose metric is given by \cite{PRL69-1849,
PRD48-1506,CQG12-2853}
\be
ds^{2}= -f(r)dt^{2} +\frac{dr^{2}}{f(r)} +r^{2}d\varphi^{2}, \label{d1}
\ee
where
\be
f(r)= -2m +\frac{r^2}{l^2} \, , \nn
\ee
in which $m$ is the mass parameter, $l$ is the AdS radius. The corresponding thermodynamic
quantities are \cite{PRD92-124069}
\be
M = \frac{r^{2}_{h}}{8l^2} \, ,~~ S = \frac{\pi}{2}r_h \, , ~~ T = \frac{r_h}{2\pi l^2} \, , ~~ V = \pi r_h^2 \, , ~~ P = \frac{1}{8\pi l^2} \, ,
\ee
where $r_{h}$ is the location of the event horizon.

It is easy to find that the Hawking temperature of the three-dimensional static neutral BTZ black hole approaches zero in the limit $r\rightarrow r_{m}$, with $r_{m}$ being the minimal event horizon radius of the black hole, but it diverges as $r\rightarrow\infty$. Consequently, the inverse temperature $\beta(r_{h})$ exhibits
\be
\beta(r_{m}) = \infty \, , \quad \beta(\infty) = 0 \label{c1}
\ee
at these asymptotic boundaries.

Using the definition of generalized off-shell Helmholtz free energy (\ref{F}) and substituting in $l^{2}=1/(8\pi P)$, we can get
\be
\mathcal{F} = \pi Pr^{2}_{h} -\frac{\pi r_h}{2\tau}.
\ee
Thus, the components of the vector $\phi$ are
\be
\phi^{r_h} = 2\pi Pr_h -\frac{\pi}{2\tau}, \qquad \phi^{\Theta} = -\cot\Theta\csc\Theta.
\ee

We now analyze the asymptotic behavior of the vector field $\phi$ near the boundary defined by Eq. (\ref{c1}). This boundary is parameterized by the closed contour $C = I_1 \cup I_2 \cup I_3 \cup I_4$,  where $I_1 = \{r_h = \infty, \Theta \in (0, \pi)\}$, $I_2 = \{r_h \in (\infty, r_m), \Theta = \pi\}$, $I_3 = \{r_h = r_m, \Theta \in (\pi, 0)\}$, and $I_4 = \{r_h \in (r_m, \infty), \Theta = 0\}$. The contour $C$ spans the entire relevant parameter space. Crucially, the construction of $\phi$ guarantees orthogonality to segments $I_2$ and $I_4$ \cite{PRL129-191101}, implying that the dominant asymptotic behavior of $\phi$ arises along $I_1$ and $I_3$. As $r_h \to r_m$ or $r_h \to \infty$, the vector $\phi$ exhibits a leftward directional shift, with its inclination governed by the value of the \(\phi^\Theta\) component. For the three-dimensional static neutral BTZ black hole, Table \ref{TableI} summarizes the directional patterns of $\phi$ on segments $I_1$ and $I_3$, alongside the corresponding global topological number $W$.

\begin{table}[t]
\caption{The direction indicated by the arrows of $\phi^{r_{h}}$ and the corresponding topological number for the static neutral BTZ black hole.}
\begin{tabular}{c|cccc|c}
\hline
Black hole solutions & $I_1$ & $I_2$ & $I_3$ & $I_4$ & $W$ \\ \hline
Static neutral BTZ black hole & $\rightarrow$ & $\uparrow$ & $\leftarrow$ & $\downarrow$ & $1$ \\
\hline
\end{tabular}
\label{TableI}
\end{table}

\begin{figure}[h]
\centering
\includegraphics[width=0.25\textwidth]{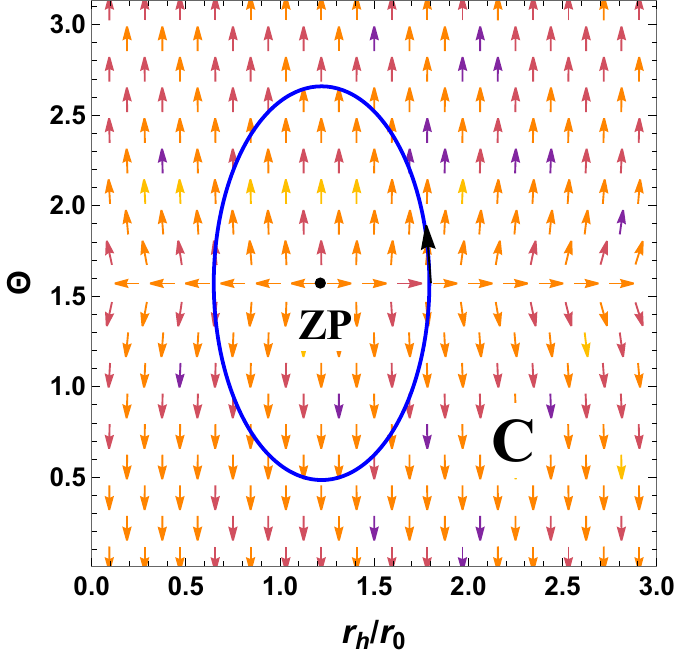}
\caption{The arrows represent the unit vector field n on a portion of the $r_{h}-\Theta$ plane with $Pr_{0}^{2}=0.005, \tau/r_{0}=40$ for the static neutral BTZ black hole. Here and hereafter, $r_0$ represents the radius of a cavity surrounding the black hole. The zero point (ZP) marked with black dot is at $(r_h/r_0,\Theta)=(1.25,\pi/2)$. The blue contour $C$ is closed loop surrounding the zero point.}\label{Fig1}
\end{figure}

\begin{figure}[h]
\centering
\includegraphics[width=0.25\textwidth]{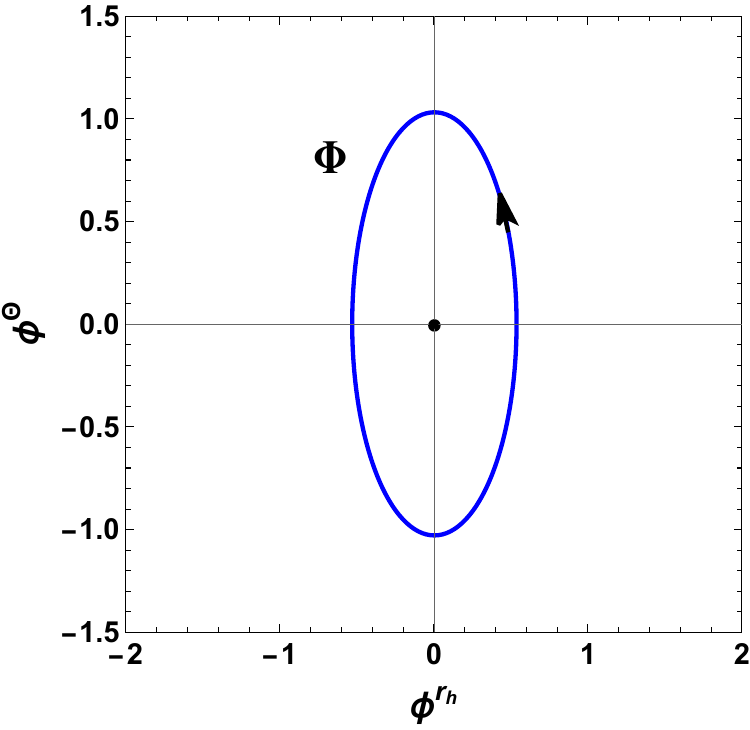}
\caption{The contour labeled $\Phi$, which represents the variation in the components of the vector field $\phi$ for the three-dimensional static neutral BTZ black hole, is denoted as $C$ in Fig.~\ref{Fig1}. The origin indicates the zero point of $\phi$. A black arrow indicates the direction of $\phi$'s rotation around the contour in Fig.~\ref{Fig1}. For contour $C$, the variation of $\phi$ causes the closed loop $\Phi$ to trace a counterclockwise path, corresponding to a winding number of $+1$.}\label{Fig2}
\end{figure}

For the three-dimensional static neutral BTZ black hole, the topological number is $W = 1$. We discuss the behavior of the components $(\phi^{r_h},\phi^{\Theta})$ of the vector $\phi$ for the static neutral BTZ black hole, analyzing each contour shown in Fig. \ref{Fig1} and illustrated in Fig. \ref{Fig2}. It is easy to find that the winding number of the zero point is $+1$.


Next, we investigate the systematic ordering of the three-dimensional static neutral BTZ black hole. The thermodynamic topology mandates that at least one black hole state exists with positive heat capacity and a local topological winding number $w = +1$. Any additional states must emerge as pairwise configurations due to the alternating sign of the heat capacity with increasing horizon radius $r_h$. Consequently, the smallest and largest black hole states in the $r_h$-ordered sequence are both thermodynamically stable. The sequence of winding numbers for all zero points follows the pattern $[+, (-, +), \ldots, +]$, where the ellipsis denotes repeated pairs of alternating charges. Crucially, the topological classification of this system is uniquely determined by the signs of the innermost and outermost winding numbers.

Then, we examine the universal thermodynamic behavior of the three-dimensional static neutral BTZ black hole. In the low-temperature limit $\beta \to \infty$, the system displays an stable small black hole state. Conversely, in the high-temperature limit $\beta \to 0$, it transitions to a stable large black hole state.

Therefore, following the thermodynamic topological classification framework proposed in Ref. \cite{PRD110-L081501}, the static neutral BTZ black hole is categorized under the class $W^{1+}$.

\section{Charged BTZ black hole}\label{III}
In this section, we investigate the universal thermodynamic topological class of the three-dimensional charged BTZ black hole \cite{PRD61-104013,PRD50-6385}. The metric is still given by Eq. (\ref{d1}), and now
\be
f(r) = -2m +\frac{r^2}{l^2} -\frac{q^2}{2}\ln\Big(\frac{r}{l}\Big) \, , \nn
\ee
where $q$ is the electric charge parameter of the black hole. In addition, the electromagnetic
gauge potential one-form is
\be
A = -q\ln\Big(\frac{r}{l}\Big)dt \, .
\ee

The thermodynamic quantities are \cite{PRD92-124069}
\be\ba
M &= \frac{r_h^2}{8l^2} -\frac{q^2}{16}\ln\Big(\frac{r_h}{l}\Big) \, , ~
S = \frac{1}{2}\pi{}r_h \, , ~ T = \frac{r_h}{2\pi l^2} -\frac{q^2}{8\pi r_h} \, , \\
\Phi &= -\frac{q}{8}\ln\Big(\frac{r_h}{l}\Big)\, , ~ Q = q , ~ V = \pi r_h^2 -\frac{\pi}{4}q^2l^2 ,  P = \frac{1}{8\pi l^2} ,
\ea\ee
where $r_h$ is the location of the event horizon.

Next, we examine the asymptotic behavior of the inverse temperature $\beta(r_{h})$ for the three-dimensional charged BTZ black hole. It is straightforward to show that the Hawking temperature diverges both as $r \to \infty$ and $r \to r_m$. Thus, it yields
\be
\beta(r_{m}) = \infty \, , \qquad  \beta(\infty) = 0  \label{c2}
\ee
for the asymptotic behaviorof the inverse temperature $\beta(r_{h})$.

According to Eq. (\ref{F}), one can compute the off-shell generalized  free energy as
\be
\mathcal{F}= \pi Pr_h^2-\frac{\pi r_h}{2\tau} -\frac{q^2}{16}\ln(2r_h\sqrt{2\pi P}) \, .
\ee
Therefore, the components of the vector $\phi$ are given by
\be
\phi^{r_h}= 2\pi Pr_h -\frac{q^2}{16r_h} -\frac{\pi}{2\tau}  \, , \quad \phi^{\Theta}=-\cot\Theta\csc\Theta \, .
\ee

We now explore the asymptotic behavior of the vector field $\phi$ near the boundaries defined by Eq. (\ref{c2}), parameterized by the contour $C = I_1 \cup I_2 \cup I_3 \cup I_4$. This contour spans the full parameter space, mirroring the methodology applied to the static neutral BTZ case.
For the three-dimensional charged BTZ black hole, Table \ref{TableII} summarizes the directional configurations of $\phi$ along segments $I_1$ and $I_3$, together with the corresponding global topological number $W$.

\begin{table}[t]
\caption{The direction indicated by the arrows of $\phi^{r_{h}}$ and the corresponding topological number for the charged BTZ black hole.}
\begin{tabular}{c|cccc|c}
\hline
Black hole solutions & $I_1$ & $I_2$ & $I_3$ & $I_4$ & $W$ \\ \hline
Charged BTZ black hole & $\rightarrow$ & $\uparrow$ & $\leftarrow$ & $\downarrow$ & $1$ \\
\hline
\end{tabular}
\label{TableII}
\end{table}

\begin{figure}[t]
\centering
\includegraphics[width=0.25\textwidth]{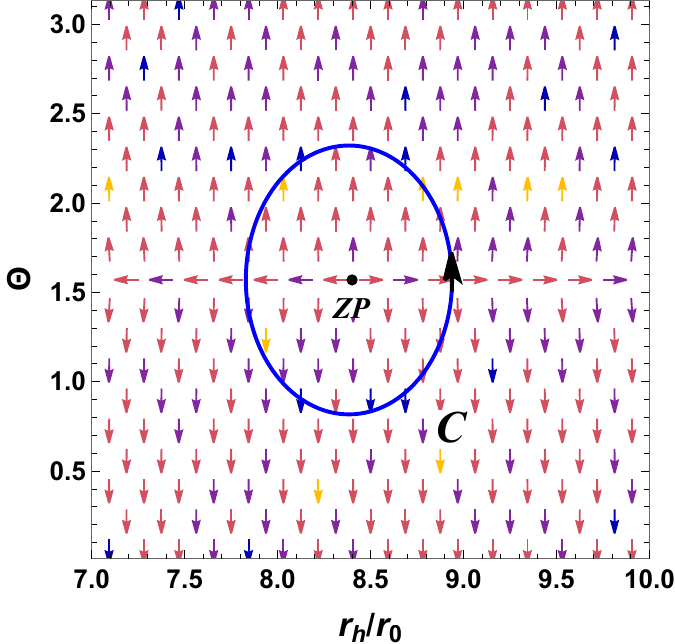}
\caption{The arrows represent the unit vector field n on a portion of the $r_{h}-\Theta$ plane with $Pr_0^2=0.001, q/r_0=1$, and $\tau/r_{0}=35$ for the charged BTZ black hole. The zero point (ZP) marked with black dot is at $(r_h/r_0,\Theta)=(8.34,\pi/2)$. The blue contour $C$ is closed loop surrounding the zero point.} \label{Fig3}
\end{figure}

\begin{figure}[t]
\centering
\includegraphics[width=0.25\textwidth]{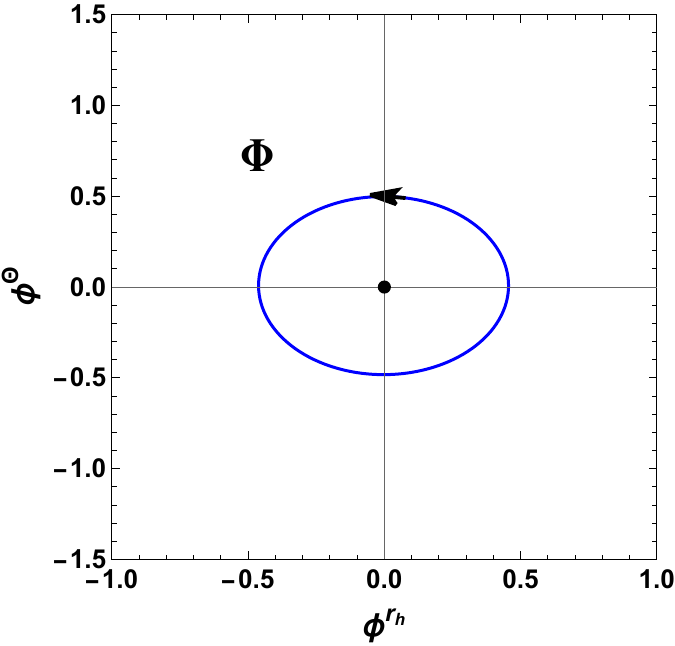}
\caption{For the three-dimensional charged BTZ black hole system, we analyze the topological contour $C$ associated with the vector field $\phi$, as shown in Fig.~\ref{Fig3}. This contour, parametrized by $\Phi$, characterizes the spatial variation of $\phi$ components. The origin of the coordinate system coincides with the zero point of $\phi$. Our visualization includes a directional black arrow indicating the rotational sense of $\phi$ along $C$. Through this analysis, we observe that the phase evolution of $\phi$ generates a counterclockwise-oriented closed loop $\Phi$, yielding a characteristic winding number $w = +1$ for contour $C$.}\label{Fig4}
\end{figure}

We analyze the behavior of the vector field components $(\phi^{r_h}, \phi^\Theta)$ for the three-dimensional charged BTZ black hole along the contours illustrated in Fig. \ref{Fig3}. The directional evolution of $\phi$ across these contours is explicitly mapped in Fig. \ref{Fig4}. It is rigorously confirmed that the zero point possesses a winding number $w = +1$.

Next, we discuss the systematic ordering of the three-dimensional charged BTZ black hole. The thermodynamic topology dictates that the three-dimensional charged BTZ black hole must contain at least one state with positive heat capacity and local topological winding number $w = +1$. Any additional states emerge as pairwise configurations due to the alternating sign of the heat capacity with increasing horizon radius $r_h$. Consequently, both the smallest and largest black hole states in the $r_h$-ordered sequence remain thermodynamically stable. The winding number sequence for all zero points follows $[+, (-, +), \ldots, +]$, where the ellipsis represents repeated pairs of alternating charges. Crucially, the system's topological classification is uniquely determined by the signs of the innermost and outermost winding numbers.

Then, we investigate the universal thermodynamic behavior of the three-dimensional charged BTZ black hole. In the low-temperature limit ($\beta \to \infty$), the system exhibits a stable small black hole state. Conversely, under high-temperature conditions ($\beta \to 0$), it transitions to a stable large black hole configuration.

In summary, according to the thermodynamic topological classification method proposed in Ref. \cite{PRD110-L081501}, the three-dimensional charged BTZ black hole is also categorized as $W^{1+}$. This classification aligns precisely with that of the three-dimensional static neutral BTZ black hole discussed in Sec.~\ref{II}. Our analysis reveals that the electric charge parameter exerts no measurable influence on the topological classification of three-dimensional static neutral BTZ black holes, a finding that stands in stark contrast to the behavior observed in four-dimensional Schwarzschild-AdS and Reissner-Nordstr\"{o}m-AdS (RN-AdS) black hole systems. This fundamental distinction constitutes one of the principal discoveries of our investigation.

\section{rotating BTZ black hole}\label{IV}
Finally, in this section, we investigate the universal thermodynamic topological class of the three-dimensional rotating BTZ black hole \cite{PRL69-1849,PRD48-1506}, whose metric reads
\be
ds^2 = -f(r)dt^2 +\frac{dr^2}{f(r)} +r^2\Big(d\varphi -\frac{J}{2r^2}dt \Big)^2 \, , \label{d3}
\ee
where
\be
f(r) = -2m +\frac{r^2}{l^2} +\frac{J^2}{4r^2} \, , \nn
\ee
in which $m$ is the mass parameter, $l$ is the AdS radius, $J$ is the angular momentum that
must satisfy $|J|\le ml$.

The thermodynamic quantities are given by \cite{PRD92-124069}
\be\ba
M &= \frac{r_h^2}{8l^2} +\frac{J^2}{32r_h^2} \, , \quad
S = \frac{\pi}{2}r_h \, , \quad \Omega = \frac{J}{16r_h^2} \, , \\
T &= \frac{r_h}{2\pi l^2} -\frac{J^2}{8\pi r_h^3} \, , \quad P = \frac{1}{8\pi l^2} \, , \quad V = \pi r_h^2 \, ,
\ea\ee
where $r_h$ is the location of the event horizon. Analysis of the three-dimensional rotating BTZ black hole reveals distinct asymptotic behaviors of the Hawking temperature. Specifically, the temperature vanishes as $r \to r_m$, while diverging in the limit $r \to \infty$. Consequently, the inverse temperature exhibits the following boundary conditions:
\be
\beta(r_m) = \infty, \quad \beta(\infty) = 0. \label{c3}
\ee

Utilizing the definition of the generalized Helmholtz free energy (\ref{F}), one can arrive at
\be
\mathcal{F} = \pi Pr^2_h -\frac{\pi r_h}{2\tau} +\frac{J^2}{32r_h^2} \, .
\ee
Therefore, the components of the vector $\phi$ are given by
\be
\phi^{r_h} = 2\pi Pr_{h} -\frac{\pi}{2\tau} -\frac{J^2}{16r^3_h}\, , \quad \phi^{\Theta} = -\cot\Theta\csc\Theta \, .
\ee

\begin{table}[t]
\caption{The direction indicated by the arrows of $\phi^{r_{h}}$ and the corresponding topological number for the rotating BTZ black hole.}
\begin{tabular}{c|cccc|c}
\hline
Black hole solutions & $I_1$ & $I_2$ & $I_3$ & $I_4$ & $W$ \\ \hline
Rotating BTZ black hole & $\rightarrow$ & $\uparrow$ & $\leftarrow$ & $\downarrow$ & $1$ \\
\hline
\end{tabular}
\label{TableIII}
\end{table}

Now, we discuss the asymptotic behavior of the vector field \(\phi\) near the boundary defined by Eq. (\ref{c3}), parameterized by the contour \(C = I_1 \cup I_2 \cup I_3 \cup I_4\). This contour spans the full parameter space, capturing the essential thermodynamic features of the system. For the three-dimensional rotating BTZ black hole, Table \ref{TableIII} summarizes the directional configurations of $\phi$ along segments $I_1$ and $I_3$, alongside the corresponding global topological number $W$.

\begin{figure}[t]
\centering
\includegraphics[width=0.25\textwidth]{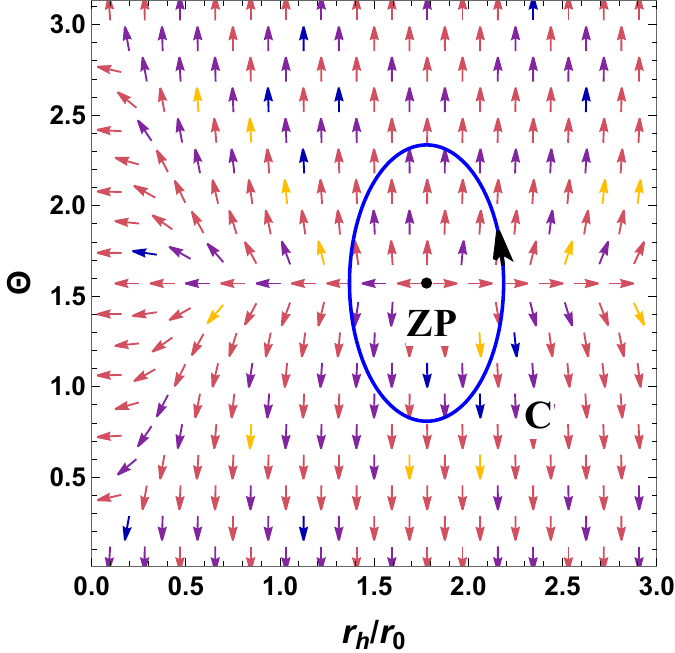}
\caption{The arrows represent the unit vector field n on a portion of the $r_{h}-\Theta$ plane with $Pr_{0}^{2}=0.01, J/r_{0}=0.6$, and $\tau/r_{0}=15$ for the rotating BTZ black hole. The zero point (ZP) marked with black dot is at $(r_{h}/r_{0},\Theta)=(1.74,\pi/2)$. The blue contour $C$ is closed loop surrounding the zero point.}\label{Fig5}
\end{figure}

\begin{figure}[t]
\centering
\includegraphics[width=0.25\textwidth]{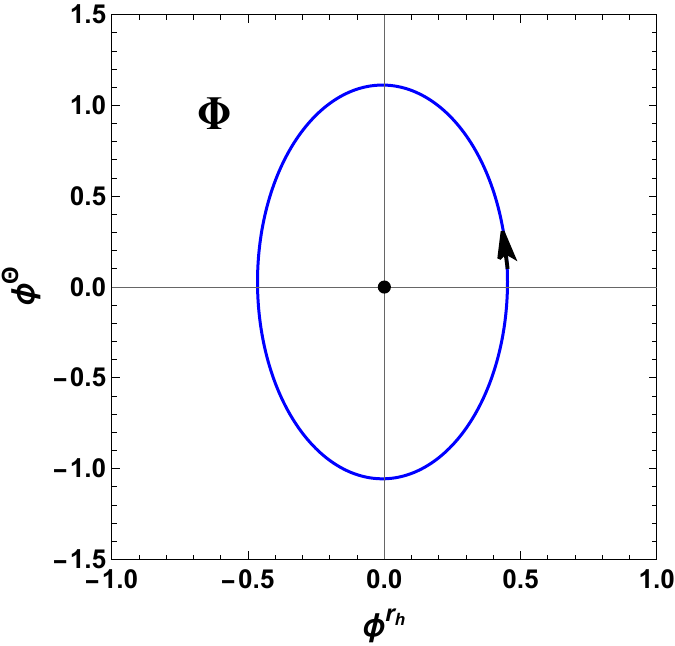}
\caption{For the three-dimensional rotating BTZ black hole, the topological contour $C$ (in Fig.~\ref{Fig5}) characterizes the $\phi$-field's structure through $\Phi$-parametrization. The vector field vanishes at the origin, while the arrow-specified rotation generates a $w=+1$ winding number via counterclockwise $\Phi$-space circulation.}\label{Fig6}
\end{figure}

For the three-dimensional rotating BTZ black hole, the topological number is $W = 1$. Consider the variations in the components $(\phi^{r_h},\phi^{\Theta})$ of the vector $\phi$ for the three-dimensional rotating BTZ black hole case along each contour depicted in Fig. \ref{Fig5}, as illustrated in Fig. \ref{Fig6}. In this case, the winding number at the only zero point is $+1$, which matches those found for the three-dimensional static neutral black hole in Sec. \ref{II}.

Next, we focus on the systematic ordering for the three-dimensional rotating BTZ black hole. The thermodynamic topology requires that there exists at least one black hole state with positive heat capacity and a local topological winding number $w = +1$. Any additional states must appear as pairwise configurations due to the alternating sign of the heat capacity as the horizon radius
$r_h$ increases. As a result, the smallest and largest black hole states in the $r_h$-ordered sequence are both thermodynamically stable. The sequence of winding numbers for all zero points follows the pattern $[+, (-, +), \ldots, +]$, where the ellipsis represents repeated pairs of alternating charges. Importantly, the topological classification of this system is uniquely determined by the signs of the innermost and outermost winding numbers.

Then, we consider the universal thermodynamic behavior of the three-dimensional rotating BTZ black hole. In the low-temperature limit, $\beta \to \infty$, the system features a small black hole that is thermodynamically stable. At high temperatures ($\beta \to 0$), the system displays an stable large black hole state.

In conclusion, following the thermodynamic topological classification method described in Ref.~\cite{PRD110-L081501}, the three-dimensional rotating BTZ black hole is classified as $W^{1+}$, belonging to the same category as both the three-dimensional static neutral BTZ black hole in Sec.~\ref{II} and the charged BTZ black hole in Sec.~\ref{III}. This demonstrates that the rotation parameter does not alter the topological classification of three-dimensional static neutral BTZ black holes, a result that stands in sharp contrast to the four-dimensional Schwarzschild-AdS and Kerr-AdS black hole cases. This constitutes another key finding of our present work.

\section{Conclusions}\label{V}
The key findings of this study are summarized in Table \ref{TableIV}. In this paper, we establish a universal thermodynamic topological classification for three-dimensional BTZ black holes, demonstrating that static neutral, charged, and rotating configurations all belong to the $W^{1+}$ category. Our analysis reveals two universal features: (I) stable small black hole states exist in the low-temperature limit for all cases, and (II) large black hole states become thermodynamically stable at high temperatures. Notably, neither electric charge nor rotation parameters affect this topological classification, in sharp contrast with their decisive role in four-dimensional black hole systems.

Our work reveals a dimensional dependence in the thermodynamic topological classification of black holes. While the thermodynamic topology of four-dimensional Schwarzschild-AdS, RN-AdS, and Kerr-AdS black holes is governed by electric charge and angular momentum parameters, their three-dimensional BTZ counterparts maintain thermodynamic topological invariance against these perturbations. This dimensional reduction effect indicates that BTZ black holes possess intrinsic topological protection mechanisms in their thermodynamic behavior, a feature absent in higher-dimensional black hole systems.

\acknowledgments

We are greatly indebted to the anonymous referee for the constructive comments to improve the presentation of this work. This work is supported by the National Natural Science Foundation of China (NSFC) under Grants No. 12205243, No. 12375053, by the Sichuan Science and Technology Program under Grant No. 2023NSFSC1347, by the Doctoral Research Initiation Project of China West Normal University under Grants No. 23KE026 and No. 21E028.

\onecolumngrid
\begin{table*}[t]
\centering
\caption{The universal thermodynamic topological classifications of the rotating black holes and their thermodynamical properties.}
\begin{tabular}{c|c|c|c|c|c|c}\hline\hline
BH solutions & Innermost & Outermost & Low $T$ & High $T$ & W & Classes \\ \hline
Static neutral BTZ BH & stable & stable & stable small  & stable large & 1 & $W^{1+}$ \\
Charged BTZ BH & stable & stable & stable small  & stable large & 1 & $W^{1+}$ \\
Rotating BTZ BH & stable & stable & stable small  & stable large & 1 & $W^{1+}$ \\
\hline\hline
\end{tabular}\label{TableIV}
\end{table*}
\twocolumngrid

\end{document}